\documentclass[12pt]{iopart}
\usepackage{mathrsfs}
\usepackage{arydshln}
\usepackage{accents}
\usepackage{color}
\usepackage{graphicx}
\usepackage{iopams}

\newtheorem{remark}{{\bf Remark}}

\begin{document}

\title[Gauge Symmetries and Noether Charges in Clebsch-Parameterized Magnetohydrodynamics]
{Gauge Symmetries and Noether Charges in Clebsch-Parameterized Magnetohydrodynamics}
\author{K Tanehashi and Z Yoshida}
\address{Graduate School of Frontier Sciences, University of Tokyo, Kashiwa, Chiba 277-8561, Japan}
\date{\today}
\ead{tanehashi@ppl.k.u-tokyo.ac.jp, yoshida@ppl.k.u-tokyo.ac.jp}

\begin{abstract}
It is shown that the Clebsch parameterization, upon canonizing the Hamiltonian system of ideal fluid or plasma dynamics, 
converts the Casimir invariants (mass and helicities) in the original noncanonical formulation into the Noether charges pertinent to the gauge symmetries of the parameterization.
The problem is addressed in the context of magnetohydrodynamics.
The concrete forms of the gauge symmetries of Clebsch parameterization are worked out. 
\end{abstract}

\pacs{47.10.Df, 45.20.Jj, 02.40.Yy, 52.55.Dy}

\submitto{\JPA}
\maketitle

\section{Introduction}
A general Hamiltonian system may have two different kinds of constants of motion (first integrals);
usual ones are caused by {symmetries} of the Hamiltonian, 
while the others, so-called \emph{Casimir invariants}, are independent of a specific choice of Hamiltonian,
but are pertinent to the underlying Poisson algebra\,\cite{ArnoldKhesin98,M98,morrison06}.
A Casimir invariant $C$ is a function (observable) such that $\{C,H \}=0$ for every $H$
(here $\{~,~\}$ denotes the Poisson bracket of a Hamiltonian system),
i.e., $C$ belongs to the center of the Poisson algebra.
The existence of a nontrivial Casimir invariant ($C\neq$ constant) is the
determining characteristics of a \emph{noncanonical} Hamiltonian system
(or, a degenerate Poisson algebra).
Conversely, a canonical Hamiltonian system does not have a Casimir invariant;
all constants of motion are of the first kind.
 
Interestingly, we may often convert a noncanonical system into a canonical system.
Then, what will become of Casimir invariants?
Here we put the method of \emph{Clebsch parameterization} into the perspective.
The noncanonical Hamiltonian system to be examined is that of the ideal magnetohydrodynamics (MHD)\,\cite{MG80}.
A Hamiltonian formulation of fluid/plasma mechanics is generally noncanonical when it is described by Eulerian variables,
and has two kinds of Casimir invariants\,\cite{M98,MG80,HKL83,H87};
one is the total mass, and the others are {helicities}.
By representing the the vector fields (fluid velocity and magnetic field) in terms of Clebsch parameters,
we can rewrite the evolution equation as a canonical Hamiltonian system, and provide it with an action principle\,
\cite{Lin,SW68,zakharov,HK83,MW83,S88,YM12,YH13}.
The aim of this work is to elucidate how the original Casimir invariants translate in the canonized system.
We will show that they become the first-kind constants of motion corresponding to the gauge symmetries of the Clebsch parameterization\,\cite{Y09}.
The gauge transformation also conserves the action integral; hence
the constants of motion are \emph{Noether charges},
i.e., the Casimir invariants translate as the Noether charges corresponding to the gauge symmetry of Clebsch parameterization.

We organize this paper as follows:
In section\,\ref{sec:formulation}, we start by reviewing the standard noncanonical Hamiltonian formulation of
MHD (subsection\,\ref{subsec:noncanonical}) and its canonization by Clebsch parameterization  (subsection\,\ref{subsec:canonical}).
In section\,\ref{sec:symmetry}, we elucidate the symmetry of the canonized system.
The Hamiltonian flow generated by the original Casimir invariant defines a symmetry group
(subsection\,\ref{subsec:symmtry_group});
the symmetry turns out to be the gauge symmetry (redundancy) of the Clebsch parameterization
(subsection\,\ref{sebsec:gauge_group}),
which yields a Noether charge that reproduces the original Casimir invariant (subsection\,\ref{subsec:Noether}).
In subsection\,\ref{subsec:generalization}, we derive an extended set of constants of motion
by studying a generalized class of symmetries.
Section \ref{sec:conclusion} concludes this work with a remark on the Lagrangian formalism and the relabeling symmetry.

\section{Hamiltonian formalisms of magnetohydrodynamics}
\label{sec:formulation}

\subsection{Ideal magnetohydrodynamics in Eulerian description}
\label{subsec:noncanonical}

The MHD system is described by
\begin{eqnarray}
\partial_{t}\rho=-\nabla\cdot(\bi{V}\rho), 
\label{MHD-1} \\
\partial_{t}\bi{V}=-(\nabla\times\bi{V})\times\bi{V}-\nabla\left(h+\frac{1}{2}V^{2}\right)+\frac{1}{\rho}(\nabla\times\bi{B})\times\bi{B}, 
\label{MHD-2} \\
\partial_{t}\bi{B}=\nabla\times(\bi{V}\times\bi{B}),
\label{MHD-3}
\end{eqnarray}
where $\rho$ is the density, $\bi{V}$ is the velocity, $\bi{B}$ is the magnetic field, and $h$ is the specific enthalpy. 
All variables are normalized by the Alfv\'en unit, i.e., the 
energy densities (thermal $h$, kinetic $\rho V^2$ and magnetic $B^2/2\mu_0$) are
normalized by the representative magnetic energy density $B_0^2/\mu_0$ ($\mu_0$ is the vacuum permeability).
Here assume a barotropic relation $h=h(\rho)$.
We consider a smoothly bounded, simply connected domain $\Omega\subset \mathbb{R}^{3}$.
We assume
\begin{eqnarray}
\bi{n}\cdot\bi{V}=0, 
\label{BC-V} \\
\bi{n}\cdot\bi{B}=0,
\label{BC-B} 
\end{eqnarray}
on the boundary $\partial\Omega$
($\bi{n}$ is the normal unit vector onto $\partial\Omega$).

We may cast the MHD equations (\ref{MHD-1})-(\ref{MHD-3}) in a Hamilton form.
A general Hamilton's equation is written as
\begin{equation}
\partial_t u=\mathcal{J} \partial_{u}H,
\label{Hamilton-general}
\end{equation}
where $u \in X$ (phase space) is a state vector,
$\mathcal{J} \in \mathrm{End}(X)$ is a Poisson operator, and $H\in C^\infty(X)$ is a Hamiltonian.
The adjoint representation of the dynamics reads
\begin{equation}
\partial_t F = \{ H, F \} ,
\label{adjoint}
\end{equation}
where $F \in C^\infty(X)$ is an observable, and $\{ A, B\} = (\partial_u A, \mathcal{J} \partial_u B )$
is the Poisson bracket induced by $\mathcal{J}$ (we denote by $( u, v )$ the inner product of  the Hilbert space $X$). 

For MHD, we define
\begin{eqnarray}
u=\left(
\begin{array}{@{\,}c@{\,}}
\rho \\ \bi{V} \\ \bi{B}
\end{array}
\right), 
\label{MHD-Hamiltonian-1}
\\
\mathcal{J}=\left(
\begin{array}{@{\,}ccc@{\,}}
0 & -\nabla\cdot & 0 \\
-\nabla & -\rho^{-1}(\nabla\times\bi{V})\times & \rho^{-1}(\nabla\times\circ)\times\bi{B} \\
0 & \nabla\times(\circ\times\rho^{-1}\bi{B}) & 0 \\
\end{array}
\right), 
\label{MHD-Hamiltonian-2}
\\
H=\int_\Omega \left(\frac{1}{2}\rho V^{2}+\rho\mathcal{E}(\rho)+\frac{1}{2}B^2\right)\rmd^{3}x, 
\label{MHD-Hamiltonian-3}
\end{eqnarray}
where $\mathcal{E}(\rho)$ the specific thermal energy,
$\partial(\rho\mathcal{E})/\partial\rho=h$ is specific enthalpy.
Evidently, substituting (\ref{MHD-Hamiltonian-1})-(\ref{MHD-Hamiltonian-3}) 
into (\ref{Hamilton-general}) reproduces the MHD equations (\ref{MHD-1})-(\ref{MHD-3}).

There are three Casimir invariants (hence, the MHD system is noncanonical):
\begin{eqnarray}
C_{1}=\int_{\Omega}\rho \rmd^{3}x, \\
C_{2}=\int_{\Omega}\bi{A}\cdot\bi{B} \rmd^{3}x, \\
C_{3}=\int_{\Omega}\bi{V}\cdot\bi{B} \rmd^{3}x,
\end{eqnarray}
where $\bi{A}=\mathrm{curl}^{-1}\bi{B}$ is the vector potential
($\mathrm{curl}^{-1}$ is the inverse operator of $\mathrm{curl}$, which is a self-adjoint operator;
see\,\cite{YoshidaGiga});
$C_{1}$ is the total mass, $C_{2}$ is the magnetic helicity, and $C_{3}$ is the cross helicity. 

\subsection{Clebsch-parameterized MHD system}
\label{subsec:canonical}
We can canonize the MHD system by parameterizing the vector fields $\bi{V}$ and $\bi{B}$
in terms of Clebsch parameters.
Here we invoke a generalized form of Clebsch parameterization such that
\begin{equation}
\bi{U}=\nabla a+\sum_{j=1}^n b_{j}\nabla c_{j}.
\label{Clebsch}
\end{equation}
When $n$ is chosen to be the dimension of base space (here, $n=3$), 
(\ref{Clebsch}) gives a complete representation of general $\bi{U}$ by
$a$, $b_{j}$ and $c_{j}$ that satisfy appropriate boundary condition independently\,\cite{Y09}.
From now on, the contraction rule on the indexes will be used to abbreviate notation.
The formulation of Clebsch-parameterized MHD will be introduced \cite{HK83,YH13}. Here we put
\begin{eqnarray}
\bi{V}=-\nabla\phi_0-\frac{\mu^\ell}{\rho}\nabla\alpha^\ell-\frac{\beta^\ell}{\rho}\nabla\phi^\ell, \\
\bi{B}=\nabla\frac{\mu^\ell}{\rho}\times\nabla\phi^\ell.
\end{eqnarray}
The index $\ell$ runs over $\{1,2,3\}$.

At $\partial\Omega$, we assume the following boundary conditions for the Clebsch parameters:
\begin{eqnarray}
\bi{n}\cdot\nabla\phi_0=0, \\
\bi{n}\cdot\nabla\alpha^\ell=0 ,\\
\frac{\mu^\ell}{\rho}=c^{\ell}, \\
\bi{n}\cdot\nabla\phi^\ell=0, \\
\bi{n}\times\nabla\frac{\beta^\ell}{\rho}=0, 
\end{eqnarray}
where $c^{\ell}$s are constant numbers, which are consistent to the boundary conditions (\ref{BC-V}) and (\ref{BC-B}) on the physical variables.
Denoting
\begin{equation}
J_c =\left(
\begin{array}{@{\,}cc@{\,}}
0 & 1 \\
-1 & 0 \\
\end{array}
\right),
\end{equation}
we define
\begin{eqnarray}
u_{c}&=&(\phi_0,\rho,\alpha^\ell,\mu^\ell,\phi^\ell,\beta^\ell)^{T},  \\
\mathcal{J}_{c}&=&\left(
\begin{array}{@{\,}ccc@{\,}}
J_c & 0 & 0 \\
0 & J_c  & 0 \\
0 & 0 & J_c  \\
\end{array}
\right), \\
H&=&\int_{\Omega}\mathcal{H}\rmd^{3}x \nonumber \\
&=&\int_{\Omega}
 \left(\frac{1}{2}\rho |\nabla\phi_0+\frac{\mu^\ell}{\rho}\nabla\alpha^\ell+\frac{\beta^\ell}{\rho}\nabla\phi^\ell|^{2}\right. \nonumber \\
&&+\left.\rho\mathcal{E}(\rho)+\frac{1}{2}|\nabla\frac{\mu^\ell}{\rho}\times\nabla\phi^\ell|^2\right)\rmd^{3}x .
\end{eqnarray}
Hamilton's equation reads as a canonical system
\begin{eqnarray}
\left\{
\begin{array}{@{\,}l}
\partial_t \phi_0=-\bi{V}\cdot\nabla\phi_0+h-V^{2}/2-\rho^{-1}\bi{J}\cdot\bi{A}, \\
\partial_t \rho=-\nabla\cdot(\rho\bi{V}), \\
\partial_t \alpha^\ell=-\bi{V}\cdot\nabla\alpha^\ell+\rho^{-1}\bi{J}\cdot\nabla\phi^\ell, \\
\partial_t \mu^\ell=-\nabla\cdot(\mu^\ell\bi{V}), \\
\partial_t \phi^\ell=-\bi{V}\cdot\nabla\phi^\ell, \\
\partial_t \beta^\ell=-\nabla\cdot(\beta^\ell\bi{V})+\bi{J}\cdot\nabla(\mu^\ell/\rho), \label{equation_of_motion} \\
\end{array}
\right.
\end{eqnarray}
where $\bi{A}=\frac{\mu^\ell}{\rho}\nabla\phi^\ell$ and $\bi{J}=\nabla\times\bi{B}$.

We have a Lagrangian (represented by the Clebsch fields)
\begin{equation}
L=\int_{\Omega}\mathcal{L}\rmd^{3}x=\int_{\Omega}\left(\rho\dot{\phi}_0+\mu^\ell\dot{\alpha}^\ell+\beta^\ell\dot{\phi}^\ell-\mathcal{H}\right)\rmd^{3}x,
\end{equation}
by which the action is written as (denoting $t=x_0$)
\begin{equation}
S=\int_{D}\mathcal{L}\rmd^{4}x,
\end{equation}
where $D$ is a space-time domain $\mathbb{R}\times\Omega$.

The invariants, $C_{1}$, $C_{2}$ and $C_{3}$ are no longer ``Casimir invariants'' in the canonized system, 
while they must be still constants of motion.
Delineation of how they are conserved in the canonized formalism is the aim of the next section.

\section{Gauge symmetries and the corresponding Noether charges}
\label{sec:symmetry}

\subsection{Symmetry groups generated by original Casimir invariants}
\label{subsec:symmtry_group}

Let $\{A,B\}_{c}=(\partial_{u_{c}}A,\mathcal{J}_{c}\partial_{u_{c}}B)$ denotes the canonical bracket.
Given a Hamiltonian $H$, a constant of motion is a functional $C$ such that
\begin{equation}
\{C,H\}_{c}=0. \label{conservation law}
\end{equation}
An infinitesimal transformation (flow) generated by a virtual Hamiltonian $G$ is given by (with small $\epsilon$)
\begin{equation}
\delta u_{G} =\epsilon\mathcal{J}_{c}\partial_{u_{c}}G.
\end{equation}
If $G$ and the actual Hamiltonian $H$ commute, i.e., if $G$ is a constant of motion,
$\delta u_{G}$ defines a \emph{symmetry group}:
(\ref{conservation law}) implies
\begin{equation}
H(u_{c}+\delta u_{G})-H(u_{c})=0.
\end{equation}
To simplify the notation, we will omit $\epsilon$ in writing perturbations $\delta u_{C}$.

The symmetry groups generated by the ``original Casimir invariants'' are 
\begin{eqnarray}
C_{1}=\int_{\Omega}\rho \rmd^{3}x\to\delta\phi_0=1 \label{total_mass_to_symmetry}, \\
C_{2}=\int_{\Omega}\bi{A}\cdot\bi{B}\rmd^{3}x\to\left\{
\begin{array}{@{\,}l}
\delta\phi_0=-\frac{2\mu^\ell}{\rho^{2}}\nabla\phi^\ell\cdot\bi{B}, \\
\delta\alpha^\ell=\frac{2}{\rho}\nabla\phi^\ell\cdot\bi{B}, \\
\delta\beta^\ell=2\nabla\frac{\mu^\ell}{\rho}\cdot\bi{B}, \\
\end{array} 
\right. \label{magnetic_helicity_to_symmetry} \\
C_{3}=\int_{\Omega}\bi{V}\cdot\bi{B}\rmd^{3}x\to\left\{
\begin{array}{@{\,}l}
\delta\phi_0=\left(\frac{\mu^\ell}{\rho^{2}}\nabla\alpha^\ell+\frac{\beta^\ell}{\rho^{2}}\nabla\phi^\ell\right)\cdot\bi{B}-\frac{\mu^\ell}{\rho^{2}}\nabla\phi^\ell\cdot\bomega, \\
\delta\alpha^\ell=-\frac{1}{\rho}\nabla\alpha^\ell\cdot\bi{B}+\frac{1}{\rho}\nabla\phi^\ell\cdot\bomega, \\
\delta\mu^\ell=-\nabla\frac{\mu^\ell}{\rho}\cdot\bi{B}, \\
\delta\phi^\ell=-\frac{1}{\rho}\nabla\phi^\ell\cdot\bi{B}, \\
\delta\beta^\ell=-\nabla\frac{\beta^\ell}{\rho}\cdot\bi{B}+\nabla\frac{\mu^\ell}{\rho}\cdot\bomega, \\
\end{array}
\right.\label{cross_helicity_to_symmetry}
\end{eqnarray}
where $\bomega=\nabla\times\bi{V}$.

\subsection{Gauge symmetries of the Clebsch parameterization}
\label{sebsec:gauge_group}
Here we show that the symmetry groups (\ref{total_mass_to_symmetry})-(\ref{cross_helicity_to_symmetry}), 
generated by the original Casimir invariants, are gauge transformations of the Clebsch parameterization,
i.e.,  the physical variables $\rho$, $\bi{V}$ and $\bi{B}$ are unchanged by the action of the group.

Under the transformation (\ref{total_mass_to_symmetry}), 
it is evident that $\delta\rho=0$ and $\delta\bi{B}=0$, 
because $\rho$ and $\bi{B}$ are independent of $\phi_0$.
Also $\delta\bi{V}=0$, because $\bi{V}$ includes $\phi_0$ in the form of its gradient.

Under (\ref{magnetic_helicity_to_symmetry}), $\delta\rho=0$ and $\delta\bi{B}=0$ are obvious,
because $\rho$ and $\bi{B}$ do not depend on $\phi_0$, $\alpha^\ell$ and $\beta^\ell$.
We also find that $\delta\bi{V}$ vanishes:
\begin{eqnarray}
\delta\bi{V} &=-\nabla\delta\phi_0-\frac{\mu^\ell}{\rho}\nabla\delta\alpha^\ell-\frac{\delta\beta^\ell}{\rho}\nabla\phi^\ell \nonumber \\
&=  \nabla\left(\frac{2\mu^\ell}{\rho^{2}}\nabla\phi^\ell\cdot\bi{B}\right)
-\frac{\mu^\ell}{\rho}\nabla\left(\frac{2}{\rho}\nabla\phi^\ell\cdot\bi{B}\right)
-\frac{1}{\rho}\left(2\nabla\frac{\mu^\ell}{\rho}\cdot\bi{B}\right)\nabla\phi^\ell
 \nonumber \\
&=\frac{2}{\rho}\left(\nabla\phi^\ell\cdot\bi{B}\right)\nabla\frac{\mu^\ell}{\rho}
-\frac{2}{\rho}\left(\nabla\frac{\mu^\ell}{\rho}\cdot\bi{B}\right)\nabla\phi^\ell
 \nonumber \\
&=\frac{2}{\rho}\bi{B}\times\left(\nabla\frac{\mu^\ell}{\rho}\times\nabla\phi^\ell\right) \nonumber \\
&=\frac{2}{\rho}\bi{B}\times\bi{B}=0.
\end{eqnarray}

Under (\ref{cross_helicity_to_symmetry}), $\delta\rho=0$ is evident.
$\delta\bi{V}$ and $\delta\bi{B}$ are calculated as follows:
\begin{eqnarray}
\delta\bi{V}&=&-\nabla\left(\left(\frac{\mu^\ell}{\rho^{2}}\nabla\alpha^\ell+\frac{\beta^\ell}{\rho^{2}}\nabla\phi^\ell\right)\cdot\bi{B}-\frac{\mu^\ell}{\rho^{2}}\nabla\phi^\ell\cdot\bomega\right) \nonumber \\
&&+\frac{1}{\rho}\left(\nabla\frac{\mu^\ell}{\rho}\cdot\bi{B}\right)\nabla\alpha^\ell \nonumber \\
&&+\frac{\mu^\ell}{\rho}\nabla\left(\frac{1}{\rho}\nabla\alpha^\ell\cdot\bi{B}-\frac{1}{\rho}\phi^\ell\cdot\bomega\right) \nonumber \\
&&+\frac{1}{\rho}\left(\nabla\frac{\beta^\ell}{\rho}\cdot\bi{B}-\nabla\frac{\mu^\ell}{\rho}\cdot\bomega\right) \nonumber \\
&&+\frac{\beta^\ell}{\rho}\nabla\left(\frac{1}{\rho}\nabla\phi^\ell\cdot\bi{B}\right) \nonumber \\
&=&\frac{1}{\rho}\left(\nabla\frac{\mu^\ell}{\rho}\cdot\bi{B}\right)\nabla\alpha^\ell-\frac{1}{\rho}\left(\nabla\alpha^\ell\cdot\bi{B}\right)\nabla\frac{\mu^\ell}{\rho} \nonumber \\
&&+\frac{1}{\rho}\left(\nabla\phi^\ell\cdot\bomega\right)\nabla\frac{\mu^\ell}{\rho}-\frac{1}{\rho}\left(\nabla\frac{\mu^\ell}{\rho}\cdot\bomega\right)\nabla\phi^\ell \nonumber \\
&&+\frac{1}{\rho}\left(\nabla\frac{\beta^\ell}{\rho}\cdot\bi{B}\right)\nabla\phi^\ell-\frac{1}{\rho}\left(\nabla\phi^\ell\cdot\bi{B}\right)\nabla\frac{\beta^\ell}{\rho} \nonumber \\
&=&\frac{1}{\rho}\bi{B}\times\left(\nabla\alpha^\ell\times\nabla\frac{\mu^\ell}{\rho}+\nabla\phi^\ell\times\nabla\frac{\beta^\ell}{\rho}\right)+\frac{1}{\rho}\bomega\times\left(\nabla\frac{\mu^\ell}{\rho}\times\nabla\phi^\ell\right) \nonumber \\
&=&\frac{1}{\rho}\left(\bi{B}\times\bomega+\bomega\times\bi{B}\right)=0, \\
\delta\bi{B}&=&\frac{\delta\mu^\ell}{\rho}\nabla\phi^\ell+\frac{\mu^\ell}{\rho}\nabla\delta\phi^\ell \nonumber \\
&=&-\nabla\times\left(\frac{1}{\rho}\left(\nabla\frac{\mu^\ell}{\rho}\cdot\bi{B}\right)\nabla\phi^\ell-\frac{1}{\rho}\left(\nabla\phi^\ell\cdot \bi{B}\right)\nabla\frac{\mu^\ell}{\rho}\right) \nonumber \\
&=&\nabla\times\left(\frac{1}{\rho}\bi{B}\times\left(\nabla\frac{\mu^\ell}{\rho}\times\nabla\phi^\ell\right)\right) \nonumber \\
&=&\nabla\times\left(\frac{1}{\rho}\bi{B}\times\bi{B}\right)=0.
\end{eqnarray}

\subsection{Noether charges}
\label{subsec:Noether}

To show that $C_{1}$, $C_{2}$ and $C_{3}$ are Noether charges of the canonized system, 
we have yet to prove that the gauge transformations (the symmetry groups) conserve the action.

The Lagrangian density $\mathcal{L}$ consists of two parts; the canonical 1-form and $\mathcal{H}$. 
Because the transformations (\ref{total_mass_to_symmetry})-(\ref{cross_helicity_to_symmetry})
are the gauge transformations,
it is obvious that $\mathcal{H}$ is unchanged.
On the other hand, the canonical 1-form may be changed by a gauge transformation. 
However, the change is an exact differential, thus the action integral is changed only by a constant.
In fact, an arbitrary Hamiltonian flow $\delta u_G$ (generated by a virtual Hamiltonian $G$)
conserves the canonical 1-form. 
Let us denote $\delta u_G/ \epsilon  = \dot{u}_c = \partial_\epsilon u_c = \mathcal{J}_{c} \partial_{u_{c}} G$.
For $p_{i}=(\rho,\mu,\beta)$ and $q^{i}=(\phi_0,\alpha,\phi)$, we observe
\begin{eqnarray}
p_{i}\dot{q}^{i}=\frac{1}{2}\partial_{\epsilon}(p_{i}q^{i})+\frac{1}{2}u_{c}\mathcal{J}_{c}^{-1}\dot{u}_c,
\end{eqnarray}
and
\begin{eqnarray}
\delta(u_{c}\mathcal{J}_{c}^{-1}\dot{u}_{c})&=&
\mathcal{J}_{c}\partial_{u_{c}}G\mathcal{J}^{-1}_{c}\dot{u}_{c}
+u_{c}\mathcal{J}_{c}^{-1}\partial_{\epsilon}(\mathcal{J}_{c}\partial_{u_{c}}G) \nonumber \\
&=&-2\dot{u}_{c}\partial_{u_{c}}G+\partial_{\epsilon}(u_{c}\partial_{u_{c}}G),
\end{eqnarray}
which implies that $\delta(p_{i}\dot{q}^{i})$ is the exact differential. 
Explicitly, the action of (\ref{magnetic_helicity_to_symmetry}) yields 
\begin{eqnarray}
\delta\mathcal{L} &= \rho\delta\dot{\phi}_0+\mu^\ell\delta\dot{\alpha}^\ell+\delta\beta^\ell\dot{\phi}^\ell \nonumber \\
&= -\partial_t\left(\bi{A}\cdot\bi{B}\right)
+\nabla\cdot\left(\frac{\mu^\ell}{\rho}\dot{\phi}^\ell\bi{B}
+\left(\dot{\bi{A}}-\nabla\left(\frac{\mu^\ell}{\rho}\dot{\phi}^\ell\right)\right)\times\bi{A}\right),
\end{eqnarray}
thus $\delta\mathcal{L}$ is an exact differential.
Similarly, by the action of (\ref{cross_helicity_to_symmetry}), we obtain
\begin{eqnarray}
\delta\mathcal{L} &=&
\rho\delta\dot{\phi}_0+\mu^\ell\delta\dot{\alpha}^\ell+\delta\mu^\ell\dot{\alpha}^\ell
+\beta^\ell\delta\dot{\phi}^\ell+\delta\beta^\ell\dot{\phi}^\ell \nonumber \\
&=&-\partial_t\left(\bi{V}'\cdot\bi{B}\right) \nonumber \\
&&+\nabla\cdot\left(-\left(\frac{\mu^\ell}{\rho}\dot{\alpha}^\ell+\frac{\beta^\ell}{\rho}\dot{\phi}^\ell\right)\bi{B}
+\left(\dot{\bi{A}}-\nabla\left(\frac{\mu^\ell}{\rho}\dot{\phi}^\ell\right)\right)\times\bi{V}'\right),
\end{eqnarray}
where $\bi{V}'=\bi{V}+\nabla\phi_0=-\mu^\ell/\rho\nabla\alpha^\ell-\beta^\ell/\rho\nabla\phi^\ell$.

Let us briefly review how a Noether charge is derived by an action-invariant transformation. 
We consider a first-order Lagrangian such that $L=\int_{\Omega}\mathcal{L}(q^{i},\partial_{\nu}q^{i})\rmd^{3}x$, which yields the Euler-Lagrange equation  $\partial_{q^{i}}\mathcal{L}-\partial_{\nu}\left(\partial_{\partial_{\nu}q^{i}}\mathcal{L}\right)=0$. 
Suppose that $L$ has a symmetry: $\delta\mathcal{L}=\partial_{\nu}\Lambda^{\nu}$ (with some vector $\Lambda^{\nu}$) 
under a transformation $q^{i} \to q^{i}+\delta q^{i}$. 
For a solution $q^{i}(t,\bi{x})$ of the Euler-Lagrange equation, we obtain
\begin{eqnarray}
\delta\mathcal{L} &= \delta q^{i}\partial_{q^{i}}\mathcal{L}+\delta(\partial_{\nu}q^{i})\partial_{\partial_{\nu}}\mathcal{L} \nonumber \\
&= \delta q^{i}\left(\partial_{q^{i}}\mathcal{L}-\partial_{\nu}\left(\partial_{\partial_{\nu}q^{i}}\mathcal{L}\right)\right)+\partial_{\nu}\left(\delta q^{i}\partial_{\partial_{\nu}q^{i}}\mathcal{L}\right) \nonumber \\
&=\partial_{\nu}\left(\delta q^{i}\partial_{\partial_{\nu}q^{i}}\mathcal{L}\right).
\end{eqnarray}
For a perturbation $\delta q^{i}$ of a symmetry group, we may write
\begin{equation}
\partial_{\nu}\left(\delta q^{i}\partial_{\partial_{\nu}q^{i}}\mathcal{L}-\Lambda^{\nu}\right)=\partial_{\nu}I^{\nu}=0,
\end{equation}
where $I^{\nu}$ is a Noether current, and $\int_{\Omega}I^{0}\rmd^{3}x=\int_{\Omega}\left(\delta q^{i}\partial_{\dot{q}^{i}}\mathcal{L}-\Lambda^{0}\right)\rmd^{3}x$ is the corresponding Noether charge.
 
Applying this procedure to the canonized MHD Lagrangian and the gauge symmetries, we obtain
\begin{eqnarray}
\delta\phi_0=1 \to\int_{\Omega}\rho \rmd^{3}x=C_{1}, \\
\left\{
\begin{array}{@{\,}l}
\delta\phi_0=-\frac{2\mu^\ell}{\rho^{2}}\nabla\phi^\ell\cdot\bi{B} \\
\delta\alpha^\ell=\frac{2}{\rho}\nabla\phi^\ell\cdot\bi{B} \\
\delta\beta^\ell=2\nabla\frac{\mu^\ell}{\rho}\cdot\bi{B} \\
\end{array} 
\right. 
\to\int_{\Omega}\bi{A}\cdot\bi{B}\rmd^{3}x=C_{2}, \\
\left\{
\begin{array}{@{\,}l}
\delta\phi_0=\left(\frac{\mu^\ell}{\rho^{2}}\nabla\alpha^\ell+\frac{\beta^\ell}{\rho^{2}}\nabla\phi^\ell\right)\cdot\bi{B}-\frac{\mu^\ell}{\rho^{2}}\nabla\phi^\ell\cdot\bomega \\
\delta\alpha^\ell=-\frac{1}{\rho}\nabla\alpha^\ell\cdot\bi{B}+\frac{1}{\rho}\nabla\phi^\ell\cdot\bomega \\
\delta\mu^\ell=-\nabla\frac{\mu^\ell}{\rho}\cdot\bi{B} \\
\delta\phi^\ell=-\frac{1}{\rho}\nabla\phi^\ell\cdot\bi{B} \\
\delta\beta^\ell=-\nabla\frac{\beta^\ell}{\rho}\cdot\bi{B}+\nabla\frac{\mu^\ell}{\rho}\cdot\bomega \\
\end{array}
\right.
\to\int_{\Omega}\bi{V}'\cdot\bi{B}\rmd^{3}x. \label{symmetry_to_cross_helicity}
\end{eqnarray}
The third Noether charge (\ref{symmetry_to_cross_helicity}) is equal to the cross helicity $C_{3}$; using the boundary condition of $\bi{B}$,
\begin{eqnarray}
\int_{\Omega}\bi{V}'\cdot\bi{B}\rmd^{3}x-\int_{\Omega}\bi{V}\cdot\bi{B}\rmd^{3}x &=\int_{\Omega}\nabla\phi_0\cdot\bi{B}\rmd^{3}x \nonumber \\
&=\int_{\Omega}\nabla\cdot\left(\phi_0\bi{B}\right) \nonumber \\
&=\int_{\partial\Omega}\bi{n}\cdot\phi_0\bi{B}\rmd^{3}x=0.
\end{eqnarray}

\begin{remark}[Gauge transformation changing the action]
A transformation conserving a Lagrangian maps a solution of the equation of motion to another solution. 
However, the converse is not necessarily true\,
In fact, there is a gauge transformation of the Clebsch parameterization which does not conserve the action.
Let $\delta\phi_0=-\frac{\phi_0\mu^\ell}{\rho^{2}}\nabla\phi^\ell\cdot\bi{B}$, $\delta\alpha^\ell=\frac{\phi_0}{\rho}\nabla\phi^\ell\cdot\bi{B}$, and $\delta\beta^\ell=\phi_0\nabla\frac{\mu^\ell}{\rho}\cdot\bi{B}$. These perturbations yield a gauge transformation:
\begin{eqnarray}
\delta\bi{V} &=-\nabla\delta\phi_0-\frac{\mu^\ell}{\rho}\nabla\delta\alpha^\ell-\frac{\delta\beta^\ell}{\rho}\nabla\phi^\ell \nonumber \\
&=\frac{\phi_0}{\rho}\left(\nabla\phi^\ell\cdot\bi{B}\right)\nabla\frac{\mu^\ell}{\rho}-\frac{\phi_0}{\rho}\left(\nabla\frac{\mu^\ell}{\rho}\cdot\bi{B}\right)\nabla\phi^\ell \nonumber \\
&=\frac{\phi_0}{\rho}\bi{B}\times\bi{B}=0,
\end{eqnarray}
It is also obvious that $\bi{B}$ and $\rho$ are unchanged. 
However, we obtain
\begin{eqnarray}
\delta\mathcal{L} &= \rho\delta\dot{\phi}_0+\mu^\ell\delta\dot{\alpha}^\ell+\delta\beta^\ell\dot{\phi}^\ell \nonumber \\
&= \phi_0\left(-\partial_t\left(\bi{A}\cdot\bi{B}\right)
+\nabla\cdot\left(\frac{\mu^\ell}{\rho}\dot{\phi}^\ell\bi{B}
+\left(\dot{\bi{A}}-\nabla\left(\frac{\mu^\ell}{\rho}\dot{\phi}^\ell\right)\right)\times\bi{A}\right)\right),
\end{eqnarray}
which is not an exact differential.
\end{remark}

\begin{figure}[h]
\begin{center}
\includegraphics[height=7.5cm]{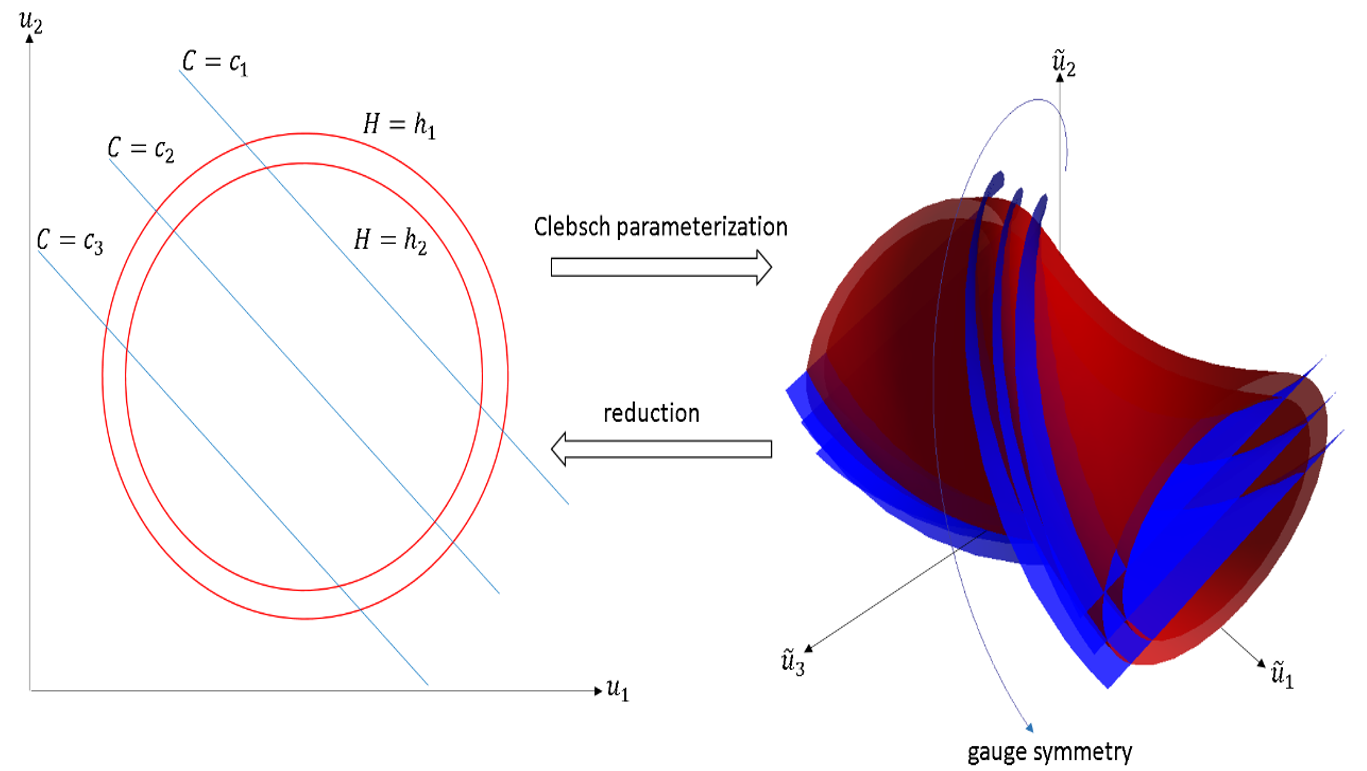}
\caption{The relation between Casimir invariants and Noether charges.
Left: In the noncanonical formulation in terms of Eulerian physical variables, 
Casimir invariants foliates the phase space.
Orbits are constrained on manifolds that are intersections of the
energy contours  (level-sets of $H$) and the Casimir leaves (level-sets of $C$).
Right: By the Clebsch parameterization, the system is embedded in a larger phase space with redundant degree of freedom;
the Hamiltonian group action generated by the Noether charges delineates the symmetry 
built in the Clebsch parameterization.}
\label{fig1}
\end{center}
\end{figure}

We have elucidated a beautiful relation, mediated by the constants $C_{1}$, $C_{2}$ and $C_{3}$, 
between the Eulerian noncanonical formalism and the Clebsch-parameterized canonical formalism; 
in the former, the constants are the Casimir invariants that foliate the phase space,
and in the latter, they are the Noether charges that identify symmetry groups
reflecting the redundancy of the Clebsch parameterization
(see Fig.\,\ref{fig1}).

\subsection{Generalized constants of motion}
\label{subsec:generalization}
The Clebsch parameterization enables us to discribe local conservation of mass and helicity. 
We define two sets of functions $G_{1}$ and $G_{2}$ such that
\begin{eqnarray}
G_{1}=\left\{a|\partial_{t}a+\bi{V}\cdot\nabla a=0 \right\}, \\
G_{2}=\left\{\lambda|\partial_{t}\lambda+\nabla\cdot\left(\lambda\bi{V}\right)=0  \right \}.
\end{eqnarray}
From (\ref{equation_of_motion}), $\phi^\ell\in G_{1}$ and $\rho, \mu^\ell\in G_{2}$. We easily verify
\begin{eqnarray}
\lambda,\eta\in G_{2}, s,t\in\mathbb{R} \to s\lambda+t\eta\in G_{2}, \label{rule1} \\
a,b\in G_{1}\to f(a,b)\in G_{1} \quad\textit{ for any smooth function } f, \label{rule2} \\
\lambda,\eta\in G_{2}\to\frac{\lambda}{\eta}\in G_{1}, \label{rule3} \\
a\in G_{1}, \lambda\in G_{2}\to a\lambda\in G_{2}, \label{rule4} \\
a,b,c\in G_{1}\to \nabla a\cdot\left(\nabla b \times\nabla c \right) \in G_{2}. \label{rule5}
\end{eqnarray}
When $\lambda\in G_{2}$, $\int_{\Omega}\lambda \rmd^{3}x$ is a constant of motion. 
From (\ref{rule3}), $\sigma^\ell=\mu^\ell/\rho\in G_{1}$ and using (\ref{rule2}) and (\ref{rule4}), $\rho f(\sigma^\ell, \phi^\ell)\in G_{2}$;
hence we find a new constant of motion  $\int_{\Omega}\rho f(\sigma^\ell,\phi^\ell)\rmd^{3}x$ ($f$ is an arbitrary smooth function).
From (\ref{rule4}) and (\ref{rule5}),${\sigma^i\nabla\phi^j\cdot\left(\nabla\sigma^k\times\nabla\sigma^\ell\right)}\in G_{2}$.
Using (\ref{rule1}), and summing up the terms such that $i=j$ and $k=l$, we obtain $\bi{A}\cdot\bi{B}\in G_{2}$.
From (\ref{rule4}), $f(\sigma^\ell,\phi^\ell)\bi{A}\cdot\bi{B}\in G_{2}$,
thus $\int_{\Omega}f(\sigma^\ell,\phi^\ell)\bi{A}\cdot\bi{B}\rmd^{3}x$ is a constant of motion.

Since $f(\sigma^\ell,\phi^\ell)$ is a scalar function that is constant along the streamlines of the flow $\bi{V}$, we may regard it as a weighting function co-moving with the plasma.
Choosing $f(\sigma^l,\phi^l)$ to be a support function on a co-moving volume element,
$\int_{\Omega}\rho f(\sigma^\ell,\phi^\ell)\rmd^{3}x$
and $\int_{\Omega}f(\sigma^\ell,\phi^\ell)\bi{A}\cdot\bi{B}\rmd^{3}x$
measure local mass and helicity contained in the volume element. These generalized constants of motion are related to the following gauge symmetries:
\begin{eqnarray}
\int_{\Omega}\rho f(\sigma^\ell,\phi^\ell)\rmd^{3}x 
\leftrightarrow\left\{
\begin{array}{l}
\delta\phi_0 =f-\frac{\mu^\ell}{\rho^{2}} f_{1}^\ell, \\
\delta\alpha^\ell=f_{1}^\ell, \\
\delta\beta^\ell=-\rho f_{2}^\ell, \\
\end{array}
\right. \label{generalized_mass_and_symmetry_MHD}\\
\int_{\Omega}f(\sigma^\ell,\phi^\ell)\bi{A}\cdot\bi{B}\rmd^{3}x \nonumber \\
\leftrightarrow\left\{
\begin{array}{l}
\delta\phi_0 =-\frac{\mu^\ell}{\rho^{2}}f_{2}^\ell\bi{A}\cdot\bi{B}-\frac{\mu^\ell}{\rho^{2}}\nabla\phi^\ell\cdot\left(f\bi{B}+\nabla\times\left(f\bi{A}\right)\right), \\
\delta\alpha^\ell=\frac{1}{\rho}f_{1}^\ell\bi{A}\cdot\bi{B}+\frac{1}{\rho}\nabla\phi^\ell\cdot\left(f\bi{B}+\nabla\times\left(f\bi{A}\right)\right), \\
\delta\beta^\ell=-f_{2}^\ell\bi{A}\cdot\bi{B}+\nabla\cdot\left(\frac{\mu^\ell}{\rho}\left(f\bi{B}+\nabla\times\left(f\bi{A}\right)\right)\right), \\
\end{array}
\right.
\end{eqnarray}
where $f_{1}^\ell$ denotes the derivative of $f$ with respect to $\sigma^\ell$, and $f_{2}^\ell$ the one with respect to $\phi^\ell$. For example, $\delta\bi{V}$ under the transformation (\ref{generalized_mass_and_symmetry_MHD}) is
\begin{eqnarray}
\delta\bi{V} &= -\nabla\left(f-\frac{\mu^\ell}{\rho^{2}}f_{1}^\ell\right)+f_{2}^\ell\nabla\phi^\ell-\frac{\mu^\ell}{\rho}\nabla f_{1}^\ell\nonumber \\
&=-\nabla f+f_{2}^\ell\nabla\phi^\ell+f_{1}^\ell\nabla\frac{\mu^\ell}{\rho} \nonumber \\
&=-\nabla f+\nabla f=0,
\end{eqnarray}
and $\delta\rho=0$ and $\delta\bi{B}=0$ are obvious,
thus (\ref{generalized_mass_and_symmetry_MHD}) is a gauge transformation.

\section{Conclusion and remarks}
\label{sec:conclusion}
We have shown, in the context of MHD, that Casimir invariants of a noncanonical Hamiltonian system 
are converted, upon canonization by an alternative parameterization (here, Clebsch parameterization) of the state vector,
into Noether charges pertinent to the gauge freedom (redundancy) of the new parameterization. 
While the conclusion delineates a simple relation between different formalisms of dynamical systems,
the specific forms of the symmetries, embodying the concrete relation, are rather complicated.
Since the Clebsch parameterization implies a nonlinear variable transformation,
the symmetries are not apparent (excepting some simple ones; cf. \cite{Y09}).
Emphasizing another aspect of the problem, we
may say that we could unearth, guided by the known Casimir invariants, 
the general gauge symmetries of the Clebsch parameterization.

There are some different methods of canonization.
A well-known method that applies for various fluid-mechanical systems
is the usage of Lagrangian variables to represent the dynamics of fluid elements;
the Lagrangian variables label the initial position of each fluid element.
Formulating an action by a Lagrangian of Newtonian-mechanical continuum,
we obtain a canonical systems of Hamilton's equation of motion\,\cite{ArnoldKhesin98,M98};
see \cite{FR60,Newcomb} for the original formulation of the Lagrangian action principle of MHD.
In the canonized system of Lagrangian variables, the Casimir invariants are translated differently.
The total mass $C_1$ and the magnetic helicity $C_2$
are now trivially conserved, because they are the integrals of the \emph{attributes} of fluid elements;
the local density and magnetic helicity are not dynamical variables, but are 
\emph{constants} bound to each fluid element (like charges of particles in particle systems).
Only the cross helicity $C_3$ is a nontrivial first integral;
it is now the Noether charge pertaining to the \emph{relabeling symmetry}
that represents the arbitrariness of providing fluid elements with Lagrangian labels\,\cite{S88,Yahalom,Padhye}.
Since Clebsch parameters may be viewed as the Eulerian counterparts of Lagrangian labels (or, the pull-back
along the Cauchy characteristics of the fluid velocity $\bi{V}$; see \cite{YM12}),
it is natural that the gauge symmetry of the Clebsch parameters (present canonization) 
and the relabeling symmetry of the Lagrangian labels (Lagrangian canonization) yield the same constant of motion.

\ack
The authors thank Dr. Yohei Kawazura for his suggestions and discussions.
This was supported by Grant-in-Aid for Scientific Research from
the Japanese Ministry of Education, Science and Culture No. 23224014.


\end{document}